\documentclass[preprint,12pt]{elsarticle}
\usepackage[utf8]{inputenc}

\biboptions{sort&compress}
\graphicspath{{images/}}
\usepackage{amsmath,amssymb,amsfonts,amsthm}

\usepackage{graphicx}

\usepackage[many]{tcolorbox}
\definecolor{revcolor}{RGB}{0,0,0} 
\newcommand{\revised}[1]{\color{black}{#1}\color{black}{}}
\newcommand{\revisedminor}[1]{\color{revcolor}{#1}\color{black}{}}

\newcommand{\customfigure}[3]{\begin{figure}[h!]%
\includegraphics[scale=#1]{#2.pdf}%
\centering%
\caption{#3}\label{fig:#2}%
\end{figure}}

\begin{document}

\begin{frontmatter}



\title{Particle coalescing with angular momentum conservation in SPH simulations}


\author{Balázs \revised{Havasi-}Tóth}

\address{Department of Hydraulic and Water Resources Engineering, Budapest University of Technology and Economics, Budapest, HUNGARY}
\address{BME-MTA Morphodynamics Research Group, Budapest, HUNGARY}

\begin{abstract}
The present work introduces a simple, yet effective particle coalescing procedure for two-dimensional SPH simulations with \revised{spatially varying} resolution. In addition to the regular conservation properties of former algorithms concerning the mass and linear momentum, the current model provides the exact conservation of the angular momentum as well. The detailed discussion of the coalescing method is followed by its verification through a frozen Taylor-Green vortex example with gradually derefined particle configuration. \revised{As a demonstration of the applicability of the proposed technique, a typical weakly compressible dam break simulation and an evolution of a two-dimensional Taylor-Green vortex pattern with local refinement are presented and comparisons \revisedminor{are made with analytical and experimental data}.}
\end{abstract}


\end{frontmatter}

\section{Introduction}
In computational science, it is highly common during the approximate solution of partial differential equations, that some requirements significantly vary in the computational domain, such as the accuracy or the level of detail of the numerical simulation results. Thus, the application of spatially varying numerical resolution plays crucial role in cost reduction of simulations, especially in computational fluid dynamics (CFD). \revised{As it has been shown by Berger and Colella in \cite{Berger1989}, mesh-based finite volume (FVM) and even finite element methods (FEM) provide robust spatial adaptivity remarkably reducing the computational costs for simulations of several applications}. Similarly, in case of particle-based Lagrangian methods adaptive resolution is also desirable, though the implementation of those models lies on slightly different assumptions.

One of the most fundamental and frequently referred meshless numerical technique is the Smoothed Particle Hydrodynamics (SPH) scheme. The first model using SPH has been constructed \revised{by Gingold and Monaghan in 1977} \citep{Gingold1977} and independently in the same year \revised{by Lucy} \citep{Lucy1977}. Initially, the method was developed for simulating astrophsysical phenomena, implying large-scale self-gravitating gaseous systems in absence of boundary conditions. Later, \revised{utilizing} the special properties of SPH, the method has been successfully applied for the simulation of both solid and fluid phases by \revised{Monaghan and Gray et al.} \citep{Monaghan1994,Gray2001}), making SPH recently an increasingly applied theory in ocean engineering and several other fields (see \citep{Chowdhury2013, Dalrymple2006, Sun2013}). Although the theory behind the variable smoothing lengths of the particles in SPH had been long since formulated by \revised{Gingold} \cite{Gingold1982} and \revised{Monaghan} \cite{Monaghan2002}, in order for the resolution in a specific region to be increased, particles need to be properly inserted and removed from the computational domain.

Based on different approaches, several techniques for spatially varying resolution have been introduced during the past two decades. Firstly, \revised{Kitsionas and Whitworth} \cite{KitsionasW2002} presented a particle splitting method for astrophysical simulations while \revised{Liu} \cite{Liu2002} achieved locally increased resolution using a Delaunay triangulation. Later, \revised{Lastiwka} \cite{Lastiwka2005} constructed criteria for local particle addition and removal with the interpolation of flow conditions and the distribution of particle mass. The precise splitting technique presented in \cite{Feldman2007} \revised{by Feldman and Bonet} became one of the most widely used methods developed futher \revised{by L{\'o}pez and Roose, Vacondio et al., Liu et al., Wang et al., Xiong et al. and Hu et al. (see \cite{Lopez2011, Vacondio2016, Liu2017, Wang2018, Xiong2013, Hu2017})} and extended for shallow water formulations by Vacondio et al. \cite{Vacondio2012, Vacondio2013b}. As a noteworthy idea, \revised{Barcarolo \cite{Barcarolo2014} and Chiron et al. \cite{Chiron2018} applied a similar splitting scheme} but without the deletion of the mother particles. Instead by turning them into passive tracers, they can be implied in the flow again later, and the daughter particles can be simply deleted when leaving the high-resolution zone.

Most recently, another approach in spatially varying resolution is based on the coupling of different fluid simulations by \revised{Bouscasse et al. and Marrone et al.} \cite{Bouscasse2013, Marrone2016} as a realization of a coupling of SPH fluid flows of different particle sizes. Such implementation is also presented \revised {by Bian et al. \cite{Bian2015} } where coupled SPH domains are overlapped through a buffer zone, in which fluid conditions are being interpolated between the zones using an overlapping interface.

According to the knowledge of the author, none of the existing particle coalescing techniques within the SPH framework provides the conservation of angular momentum. Regardless of the coarsening procedure, the aforementioned techniques reduce the degrees of freedom of the original configuration so that the angular momentum inevitably vanishes in the merged or derefined configuration. In the present work, a particle coalescing technique is introduced, which exaclty preserves not only the mass and the the linear momentum, but the angular momentum as well without significant additional computational cost compared to the existing methods.

\section{Motivation}
The particle merging process introduced by \revised{Vacondio et al.} \cite{Vacondio2013a} performs coalescing of particle pairs in such a way that a candidate particle $a$ and its closest neighbor $b$ are replaced with a new particle $m$ in the common center of mass of particles $a$ and $b$. As it has been shown in \cite{Vacondio2013a}, pairwise coalescing process can be formulated so that mass and linear momentum are exactly preserved. However, during the merge, the degrees of freedom of the local particle set $\{a,b\}$ and $\{m\}$ is reduced from $4$ to $2$ in two dimensions. As a result, in case of the velocity vectors of the original pair of particles are not equal ($\textbf{v}_a\neq\textbf{v}_b$), the merge leads to a complete loss of angular momentum.
\customfigure{0.3}{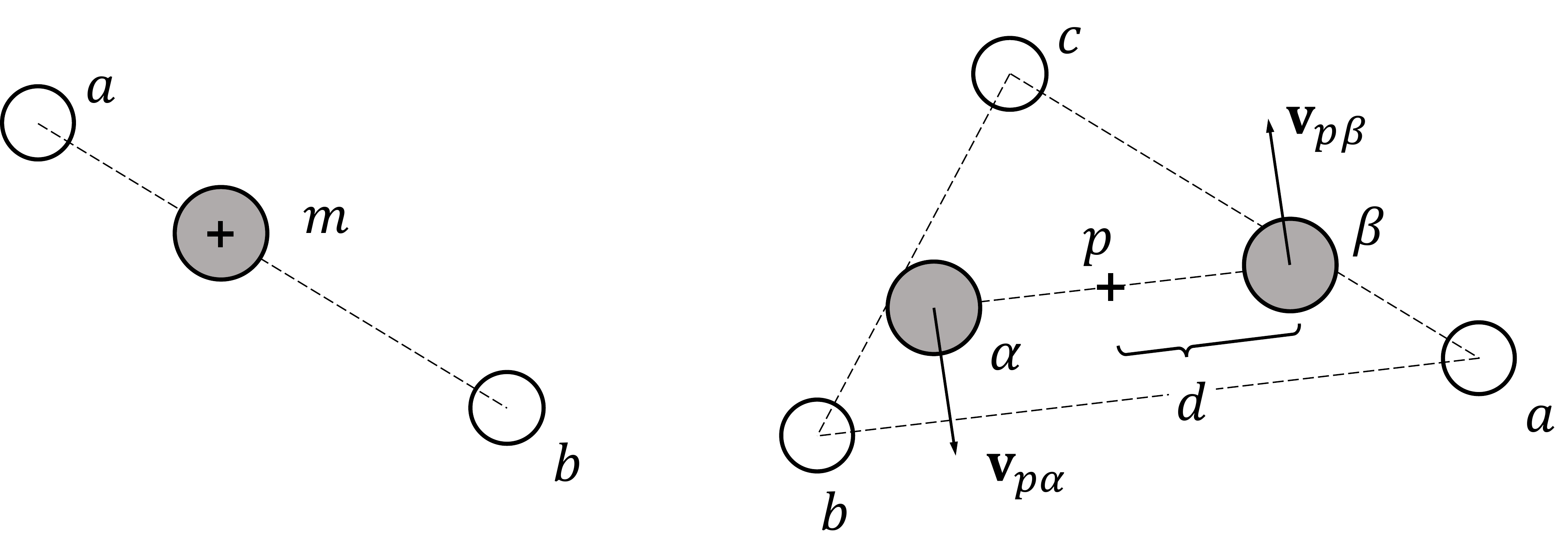}{Coalescing of a particle pair (left) and a triplet (right). The original and the new particle layout are shown by the empty and shaded circles respectively.}
The basic idea behind the proposed resolution reduction technique is that in two dimensions the angular momentum can be preserved only if the remaining degrees of freedom after merging is at least $4$. Thus, in the present work we introduce a modified particle coalescing technique in which the particle merging is performed over particle triplets instead of pairs. The difference between the two methods is shown in Figure \ref{fig:layout}. After merging, a triplet $\{a,b,c\}$ is replaced with a pair of identical particles $\{\alpha,\beta\}$, reducing the initial degrees of freedom from $6$ to $4$, which is still sufficient to preserve both the local linear and angular momenta of the original layout.

\section{The merging procedure}
\revised{Besides the criteria of the pre-defined refinement zones, in the present work, a particle is chosen to be a candidate for coalescing when its mass does not exceed the 90\% of the mass corresponding to the original coarse resolution. Also, similarly to the process proposed by \cite{Vacondio2013a}, the triplet is formed using the two closest neighbors of the candidate particle. To avoid a particle to be selected for two candidates in the same time step, the selection process excludes all particles that are already marked for coalescing. The merging of the triplets is performed only after all candidates have been processed -- either found neighbors or not.}
\subsection{Conservation of mass and linear momentum}
Before placing the new particles, the position and velocity of the center of mass of the particles to be merged need to be computed. Furthermore, by keeping these quantities the same before and after the merge, the linear momentum is also preserved. As a result, we require, that:
\begin{equation}
\begin{split}
&\textbf{r}_p=\frac{\textbf{r}_a m_a+\textbf{r}_b m_b+\textbf{r}_c m_c}{m_a+m_b+m_c}=\frac{\textbf{r}_\alpha m_\alpha+\textbf{r}_\beta m_\beta}{m_\alpha+m_\beta},\\
&\textbf{v}_p=\frac{\textbf{v}_a m_a+\textbf{v}_b m_b+\textbf{v}_c m_c}{m_a+m_b+m_c}=\frac{\textbf{v}_\alpha m_\alpha+\textbf{v}_\beta m_\beta}{m_\alpha+m_\beta}.
\end{split}
\end{equation}
The exact conservation of mass is also enforced by choosing the new masses as
\begin{equation}
m_\alpha=m_\beta=m_m=\frac{m_a+m_b+m_c}{2}.
\end{equation}
\subsection{Computation of the smoothing length and the new positions}
{As it has been pointed out \revised{by Feldman and Bonet} \cite{Feldman2007},} to minimize the influence of the change in the particle layout on the solution, the minimization of the density variation during particle splitting and coalescing is of crucial importance. In attempt to minimize the change in the density field, we compute the density at $\textbf{r}_p$ for both the original and coalesced layout, which therefore should be equal:
\begin{equation} \label{eq:density_eval}
\langle\rho_p\rangle=\sum_{i=a,b,c}{m_i W(\vert\textbf{r}_p-\textbf{r}_i\vert,h_i)}=(m_\alpha+m_\beta)(W(\vert\textbf{r}_p-\textbf{r}_\alpha\vert,h_\alpha))=(m_\alpha+m_\beta)\revisedminor{W_{p\alpha}},
\end{equation}
where for the sake of simplicity of the subsequent assumptions, we consider the two-dimensional Gaussian smoothing kernel function
\begin{equation} \label{eq:gaussian}
W(r,h) = \frac{1}{\pi h^2}\exp{\bigg(-\frac{r^2}{h^2}\bigg)}.
\end{equation}
In contrast with \cite{Vacondio2013a}, the smoothing radius cannot be computed directly due to $r\neq0$ takes place in (\ref{eq:gaussian}). Furthermore, as it is visualized in Figure \ref{fig:envelope}, normalized kernels cannot produce arbitrary high values at a given place by varying the smoothing radius, therefore the distance $d=\vert\textbf{r}_\alpha-\textbf{r}_\beta\vert/2$ should be chosen carefully to avoid infeasible solutions of (\ref{eq:density_eval}).
\customfigure{0.4}{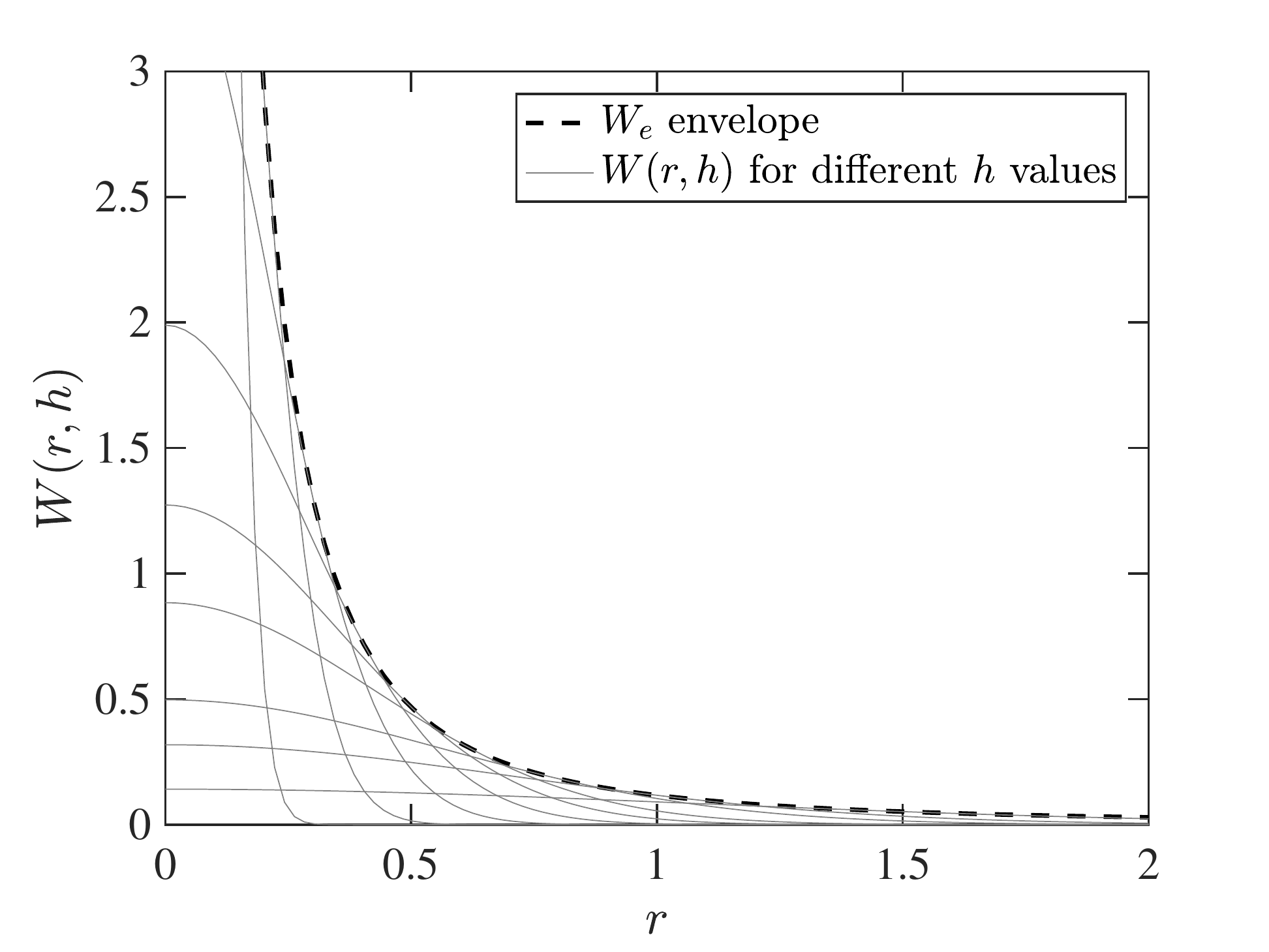}{Gaussian kernel functions with different smoothing radii (gray solid curves) and the envelope (black dashed curve) marking the maxima of $x$ for the corresponding values of $W(x,h)$.}
To compute the maximum distance $r_{max}=\vert\textbf{r}_\alpha-\textbf{r}_p\vert=\vert\textbf{r}_\beta-\textbf{r}_p\vert$, at which (\ref{eq:density_eval}) can be satisfied, (\ref{eq:gaussian}) has been analytically analysed. Firstly, using the known value of $\revisedminor{W_{p\alpha}}$ from (\ref{eq:density_eval}), consider the inverse of the Gaussian kernel function:
\begin{equation} \label{eq:inv_kernel}
\revisedminor{r=h\sqrt{-\ln(W_{p\alpha}\pi h^2)}}.
\end{equation}
To find its maxima, the derivative with repsect to the smoothing length is computed and constrained to zero as follows
\begin{equation} \label{eq:inverse_derivative}
\frac{\partial r}{\partial h}=-\frac{\ln(\revisedminor{W_{p\alpha}}\pi h^2)+1}{\sqrt{\revisedminor{-}\ln(\revisedminor{W_{p\alpha}}\pi h^2)}}=0.
\end{equation}
This equation has finite roots only if the numerator is zero, thus the solution of
\begin{equation}
\ln(\revisedminor{W_{p\alpha}}\pi h^2)=-1
\end{equation}
gives
\begin{equation}
h_{max}=\revisedminor{\pm}\sqrt{\frac{1}{e\pi \revisedminor{W_{p\alpha}}}},
\end{equation}
\revisedminor{from which the positive value is} the smoothing length corresponding to the maximum value of the distance $r_{max}$ for a given value of $W$.
Finally, substituting $h_{max}$ to (\ref{eq:inv_kernel}) gives
\begin{equation} \label{eq:maxdist}
r_{max}=\sqrt{-\frac{1}{e\pi \revisedminor{W_{p\alpha}}}\ln\bigg(\frac{\revisedminor{W_{p\alpha}}\pi}{e\pi \revisedminor{W_{p\alpha}}}\bigg)}=\sqrt{\frac{1}{e\pi \revisedminor{W_{p\alpha}}}}=h_{max},
\end{equation}
which yields that $x=h$ is a nullcline of (\ref{eq:inv_kernel}) for arbitrary $W$, shown in Figure \ref{fig:nullcline}.
From (\ref{eq:maxdist}), the envelope Figure \ref{fig:envelope} of the Gaussian kernel can be expressed as
\begin{equation}
W_e = \frac{1}{e\pi x^2}.
\end{equation}

\customfigure{0.4}{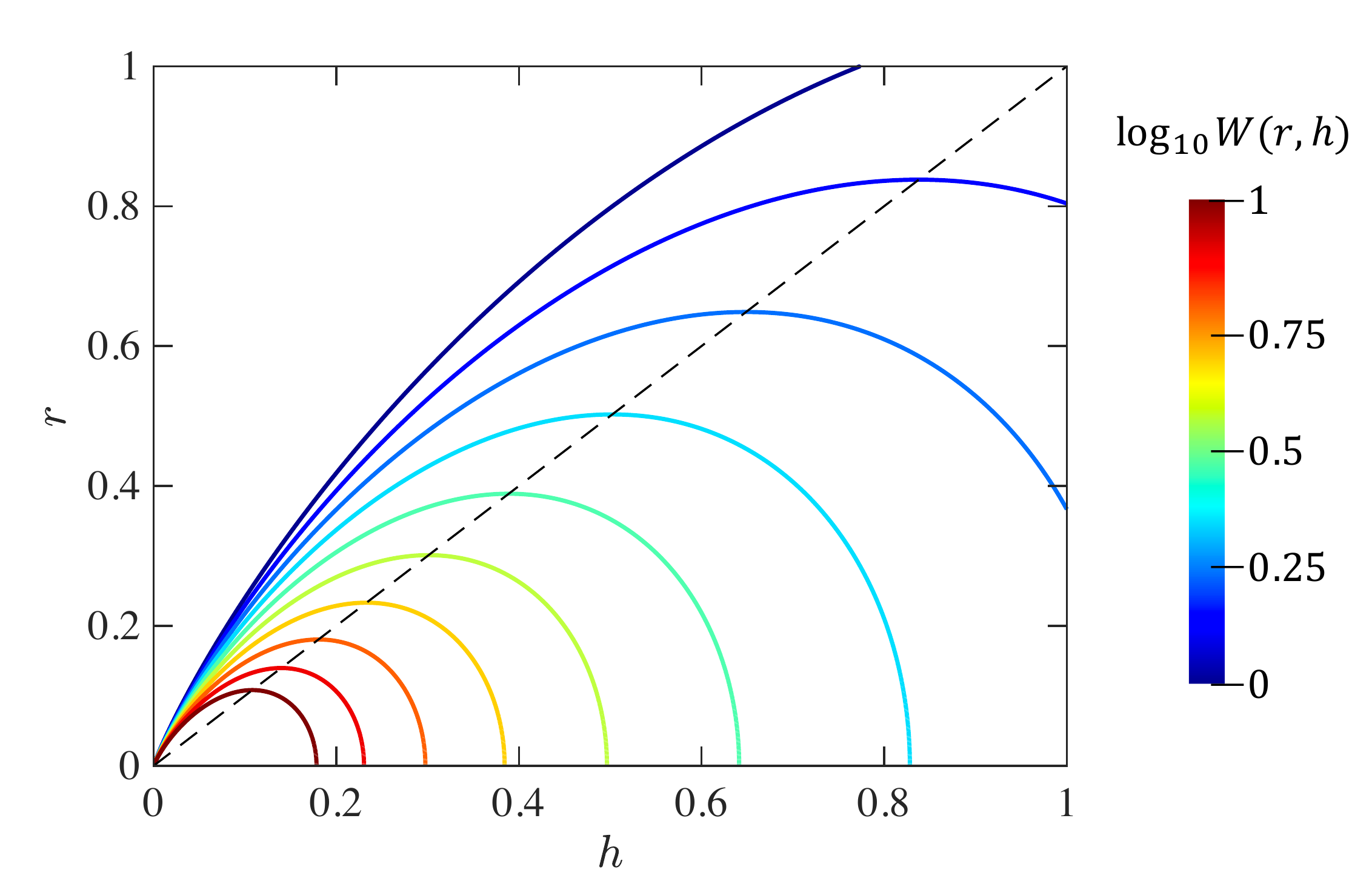}{Isocurves of the Gaussian kernel on the $r-h$ plane (solid lines), and the $r=h$ nullcline (dashed line).}

Although the maximum distance $r_{max}=\vert\textbf{r}_\alpha-\textbf{r}_p\vert$ can be computed from the simple expression (\ref{eq:maxdist}), the optimal distance is usually smaller. In the present work, the new spacing was chosen to be equal to the average distance of the original particles from $\textbf{r}_p$:
\begin{equation}
d=\eta\frac{1}{3}\sum_{i=a,b,c}{\vert\textbf{r}_i-\textbf{r}_p\vert},
\end{equation}
where $\eta$ is a scaling constant, which results a uniform particle distribution when set between $0.9$ and $1$, and $d$ limited to $r_{max}$ only if $d\geq r_{max}$. Otherwise the smoothing radius $h_\alpha=h_\beta$ needs to be computed. Since (\ref{eq:gaussian}) cannot be solved directly for $h$, the iterative solution
\begin{equation}
h^{n+1}=\sqrt{\frac{1}{\pi \revisedminor{W_{p\alpha}}}\exp\bigg(-\frac{r^2}{(h^n)^{2}}\bigg)}
\end{equation}
is used, where the superscript denotes the iteration level and $h^0=h_{max}$. The iteration is terminated \revised{when} the
\begin{equation}
\revisedminor{10^{-6}>\bigg\vert\frac{h^{n+1}-h^{n}}{h^{n}}\bigg\vert}
\end{equation}
exiting criterion is satisfied.

After computing the distance $d$ and smoothing radius $h$, the positions $\textbf{r}_\alpha$ and $\textbf{r}_\beta$ are obtained using the simple constraint
\begin{equation}
\textbf{r}_{\alpha}-\textbf{r}_{\beta}\quad\big\vert\big\vert\quad \max(\textbf{r}_{ab},\textbf{r}_{ac},\textbf{r}_{bc}).
\end{equation}

\subsection{Conservation of angular momentum}
The linear momentum is exactly preserved if the
\begin{equation} \label{eq:glob_lin_momentum}
\sum_{i=a,b,c}{m_i\textbf{v}_i}=m_\alpha\textbf{v}_\alpha+m_\beta\textbf{v}_\beta=m_m(\textbf{v}_\alpha+\textbf{v}_\beta)=(m_\alpha+m_\beta)\textbf{v}_p\\
\end{equation}
condition is satisfied. Considering the center of mass instead of the origin, (\ref{eq:glob_lin_momentum}) can be written in relative coordinates, implying that the linear momentum vanishes in the co-moving frame:
\begin{equation} \label{eq:rel_lin_momentum}
\sum_{i=a,b,c}{m_i\textbf{v}_{pi}}=m_\alpha\textbf{v}_{p\alpha}+m_\beta\textbf{v}_{p\beta}=m_m(\textbf{v}_{p\alpha}+\textbf{v}_{p\beta})=0,\\
\end{equation}
which in turn yields to
\begin{equation} \label{eq:lin_velocity}
\textbf{v}_{p\alpha}=-\textbf{v}_{p\beta},
\end{equation}
where $\textbf{v}_{p\alpha}=\textbf{v}_{p}-\textbf{v}_{\alpha}$. Thus, compared to \cite{Vacondio2013a}, the degrees of freedom additionally implied in our model can be \revised{utilized} to carry non-zero velocities without violating the linear momentum conservation.

Similarly to (\ref{eq:rel_lin_momentum}), the angular momentum conservation can be expressed in the co-moving frame. Considering that the particle layout and merging process is being performed in two dimensions, the angular momentum becomes a scalar quantity. 
Using (\ref{eq:lin_velocity}), the angular momentum in terms of the local center of mass reads as
\begin{equation}
\sum_{i=a,b,c}{m_i(\textbf{r}_{ip}\times\textbf{v}_{ip})\textbf{e}_k}=(m_\alpha\textbf{r}_{\alpha p}\times\textbf{v}_{\alpha p}+m_\beta\textbf{r}_{\beta p}\times\textbf{v}_{\beta p})\textbf{e}_k=(m_\alpha+m_\beta)(\textbf{r}_{\alpha p}\times\textbf{v}_{\alpha p})\textbf{e}_k,
\end{equation}
which can be directly applied to compute the velocities $\textbf{v}_{\alpha p}$ and $\textbf{v}_{\beta p}$, hence the velocities in the global frame can be expressed as
\begin{equation}
\begin{split}
\textbf{v}_\alpha=\textbf{v}_{\alpha p}+\textbf{v}_p, \\
\textbf{v}_\beta=\textbf{v}_{\beta p}+\textbf{v}_p,
\end{split}
\end{equation}
\revised{where the $\textbf{v}_p$ velocity of the local center of mass is already known from the unmodified configuration.}

\section{Fluid flow modelling with SPH}
This section briefly presents the most fundamental and conventional fluid flow model of the Lagrangian particle-based SPH method applied in the present work.
\subsection{Governing equations}
Considering the substantial or material time-derivative of any quantity $A$ as the sum of the local and convective terms
\begin{equation}
\frac{dA}{dt}=\frac{\partial A}{\partial t}+v\cdot\nabla A,
\end{equation}
the governing equations of an arbitrary, compressible inviscid fluid can be written in the Lagrangian frame \revised{\cite{Monaghan2005}}:
\begin{equation} \label{eq:fluid_equations}
\begin{split}
&\frac{d\rho}{dt}=-\rho\nabla \textbf{v}, \\
&\frac{d\textbf{v}}{dt}=-\frac{1}{\rho}\nabla p+\textbf{g}+\frac{\textbf{f}}{\rho},
\end{split}
\end{equation}
where $\rho$, $p$, $\textbf{v}$ and $\textbf{f}$ are the density, pressure, velocity and the sum of the specific external forces respectively and $\textbf{g}$ is the gravitational acceleration vector. \revised{As proposed by Monaghan \cite{Monaghan2005}, the weakly compressible variant of SPH is used in the present work to control compressibility with the equation of state}
\begin{equation} \label{eq:EOS}
p=c^2(\rho-\rho_0),
\end{equation}
where $c$ is the artificially reduced speed of sound and $\rho_0$ is the reference density of the fluid. The speed of sound is usually tuned to keep the maximum density variations around $1\%$ using the following condition:
\begin{equation}
c=10v_{max},
\end{equation}
where $v_{max}$ is the expected maximum of the velocity magnitude in the specific flow problem.
\subsection{The SPH discretization}
SPH is a pure Lagrangian meshless collocation method for the solution of time-dependent partial differential equations (PDEs) such as (\ref{eq:fluid_equations}). The fundamental idea is that any function $A$ can be reconstructed in the domain $\Omega$ using the generalized interpolation also known as convolution \cite{Monaghan2005}
\begin{equation} \label{eq:continuum_convolution}
\langle A(\textbf{r})\rangle=\int_\Omega{A(\textbf{r}-\textbf{r}')\delta(\textbf{r}-\textbf{r}')d\textbf{r}},
\end{equation}
which, after \revisedminor{substituting} the Dirac-function $\delta(\textbf{r}-\textbf{r}')$ \revisedminor{with} an appropriate mollifier $W_{ij}=W(\vert\textbf{r}_i-\textbf{r}_j\vert)$, can be written for finite-size nodes (particles) in discrete form as
\begin{equation} \label{eq:discrete_convolution}
\langle A(\textbf{r}_i)\rangle=\langle A_i\rangle=\sum_j{A_j W_{ij}\frac{m_j}{\rho_j}},
\end{equation}
where $m_j$ and $\rho_j$ are the mass and density assigned to particle $j$. \revised{Among several other works, a detailed introduction of the discretization procedure is given in \cite{Violeau2012} by Violeau, as well as in the review of the SPH scheme by \cite{Monaghan2005}.}

The expression (\ref{eq:discrete_convolution}) can be used for the construction of the algebraic approximation of local spatial derivatives such as gradient and divergence of any function. Thus, \revised{choosing the proper differential operators -- concerning conservativity and consistency properties discussed by Sigalotti et al. \cite{Sigalotti2016} and Vaughan \cite{Vaughan2008} --} the divergence of the velocity field and the pressure gradient can be written as
\begin{equation} \label{eq:sph_diffop}
\begin{split}
&\langle\nabla\cdot\textbf{v}_i\rangle=\sum_j(\textbf{v}_j-\textbf{v}_i)\nabla W_{ij}\frac{m_j}{\rho_j}, \\
&\langle\nabla p_i\rangle=\rho_i\sum_j\bigg(\frac{p_i}{\rho_i^2}+\frac{p_j}{\rho_j^2}\bigg)\nabla W_{ij}m_j.
\end{split}
\end{equation}

Through the conventional SPH discretization process, the PDE system (\ref{eq:fluid_equations}) is converted into a set of ordinary differential equations (ODE's) by substituting the spatial derivatives with the SPH algebraic representations. Thus, the conventional SPH fluid-equations read as
\begin{equation} \label{eq:fluid_equations_discretized}
\begin{split}
&\frac{d\rho_i}{dt}=-\rho_i\sum{(\textbf{v}_j-\textbf{v}_i)\frac{m_j}{\rho_j}\nabla W_{ij}}, \\
&\frac{d\textbf{v}_i}{dt}=-\sum_j{\bigg(\frac{p_i}{\rho_i^2}+\frac{p_j}{\rho_j^2}\bigg)m_j\nabla W_{ij}}+\textbf{g}+\frac{\textbf{f}_i}{\rho_i}, \\
\end{split}
\end{equation}
where the subscripts denote particle $i$ and its neighbouring particles $j$ and $W_{ij}=W(\textbf{r}_i-\textbf{r}_j,h)$ is chosen to be the fifth order polynomial Wendland smoothing kernel function \revised{\cite{Wendland1995}}:
\begin{equation} \label{eq:kernel_wendland}
W(q)=
\begin{cases}
C_d\big(1-\frac{q}{2}\big)^4(2q+1) \quad \rm{if} \quad q\leq2 \\
0 \quad \rm{if} \quad q>2
\end{cases},
\end{equation}
where $q=\vert \textbf{r}_{ij}\vert/h$, $C_d=7/(4\pi h^2)$ is the normalization constant in two dimensions and $h$ is the smoothing radius. In the present work, the interparticle distance is chosen to be $dx=2.1h$.

To facilitate numerical stability and reduce spurious oscillations in the solution, the equations (\ref{eq:fluid_equations_discretized}) are usually extended by artificial diffusive terms \revised{proposed by Morris et al. \cite{Morris1997} as well as Molteni and Colagrossi \cite{Molteni2009}}. Here, the stabilization terms presented \revised{by Antuono et al.} \cite{Antuono2010} are applied for both the continuity and momentum equations, hence the whole system reads as:
\begin{equation} \label{eq:whole_system}
\begin{split}
&\frac{d\rho_i}{dt}=-\rho_i\sum{(\textbf{v}_j-\textbf{v}_i)\frac{m_j}{\rho_j}\nabla W_{ij}}+\xi hc\sum_j\Psi_{ij}\nabla W_{ij}\frac{m_j}{\rho_j}, \\
&\frac{d\textbf{v}_i}{dt}=-\sum_j{\bigg(\frac{p_i}{\rho_i^2}+\frac{p_j}{\rho_j^2}\bigg)m_j\nabla W_{ij}}+\textbf{g}+\frac{\textbf{f}_i}{\rho_i}+\alpha hc\rho_0\sum_j\pi_{ij}\nabla W_{ij}\frac{m_j}{\rho_j}, \\
&p_i=c^2(\rho_i-\rho_0),
\end{split}
\end{equation}
where $\xi=0.1$ and $\alpha=0.01$ are diffusion constants\revised{ and
\begin{equation}
\begin{split}
&\Psi_{ij}=2(\rho_j-\rho_i)\frac{\textbf{r}_{ji}}{\vert\textbf{r}_{ji}\vert^2}\revisedminor{-}\big[\langle \nabla\rho \rangle^L_i+\langle \nabla\rho \rangle^L_j\big], \\
&\pi_{ij}=(u_j-u_i)\frac{\textbf{r}_{ji}}{\vert\textbf{r}_{ji}\vert^2},
\end{split}
\end{equation}
where $\langle \nabla . \rangle^L$ is the renormalized \revisedminor{gradient formula introduced by Randles and Libersky \cite{Randles1996}} applied here for the density to fix the violation of mass conservation caused by the kernel truncation near the free surfaces}. \revised{However, in cases when the system (\ref{eq:fluid_equations}) implies kinematic viscosity $\nu\Delta\textbf{v}$ in the momentum equation, the corresponding artificial diffusive term can be changed to form the final system
\begin{equation}
\frac{d\textbf{v}_i}{dt}=-\sum_j{\bigg(\frac{p_i}{\rho_i^2}+\frac{p_j}{\rho_j^2}\bigg)m_j\nabla W_{ij}}+\textbf{g}+\frac{\textbf{f}_i}{\rho_i}+2\nu\sum_j\frac{(\textbf{v}_i-\textbf{v}_j)\textbf{r}_{ij}}{\vert\textbf{r}_{ij}\vert^2}\nabla W_{ij}\frac{m_j}{\rho_j},
\end{equation}
where $\nu$ denotes the strength of the viscosity.
}

\revised{In case of zones indicating low pressure regions in SPH flow simulations, non-physical voids may be formed in the particle distribution due to the tensile instability problem. To fill the undesired cavities in the fluid, the $\delta^+$-SPH method is recently proposed by Sun et al. \cite{Sun2018} applying a modified particle shifting technique (PST) that implies the tensile instability correction introduced by Monaghan \cite{Monaghan2000}. Also, particles may form unisotropic linear structures in the vicinity of flattening motion leading to a rapid decay of accuracy. This behavior is usually avoided by using the PST as well. The shifting rate of the particles is computed in every time step as
\begin{equation}
\begin{split}
\textbf{r}^*_i&=\textbf{r}_i+\delta \textbf{r}_i, \\
\delta \textbf{r}_i&=-CFL\frac{v_{max}}{c}(2h)^2\sum_j{\Bigg[1+R\Bigg(\frac{W_{ij}}{W(\Delta x)}\Bigg)^n\Bigg]\nabla W_{ij}\frac{m_j}{\rho_i+\rho_j}},
\end{split}
\end{equation}
where $\textbf{r}^*$ is the new particle position, $CFL=1.5$ is the Courant-Friedrichs-Levy number, $\Delta x$ is the average interparticle distance, $n=4$ and $R=0.2$.
}
\section{Results and discussion}
\subsection{Frozen Taylor-Green vortex}
A simple case for verification of the angular momentum conservation is the investigatation of a single, high resolution vortex under gradual derefinement without the presence of physical or artificial dissipation. For the initial layout, we constructed a $209\times209$ grid of particles in an axis-aligned origin-centered unit square with the velocity field given by the velocity field of the Taylor-Green vortex sheet
\begin{equation}
\textbf{v}_0=\begin{bmatrix}
           sin(\pi(x-0.5))cos(\pi(y-0.5)) \\
           -cos(\pi(x-0.5))sin(\pi(y-0.5))
         \end{bmatrix}.
\end{equation}
Without moving the Lagrangian particles in space due to their velocities, the initial layout has been gradually derefined until the number of particles reached the order of magnitude of $10$, which is considered to be a coarse representation of a vortex. As Figure \ref{fig:resolution} shows, similarly to the conventional particle coalescing techniques, the uniformity of the particle layout is preserved during the derefinement process. In turn, presented in Figure \ref{fig:decay}, the merging process based on the particle triplets preserves the angular momentum exactly. In comparison with pairwise coalescing techniques, our method reduces the particle number in every derefinement step with the theoretical maximum rate of $33.\dot 3\%$ instead of $50\%$. As it is visible in Figure \ref{fig:decay}, this moderate reduction rate causes the increment of the required steps (from $20$ to $30$ in this specific case) until reaching the desired coarse resolution.
\customfigure{0.45}{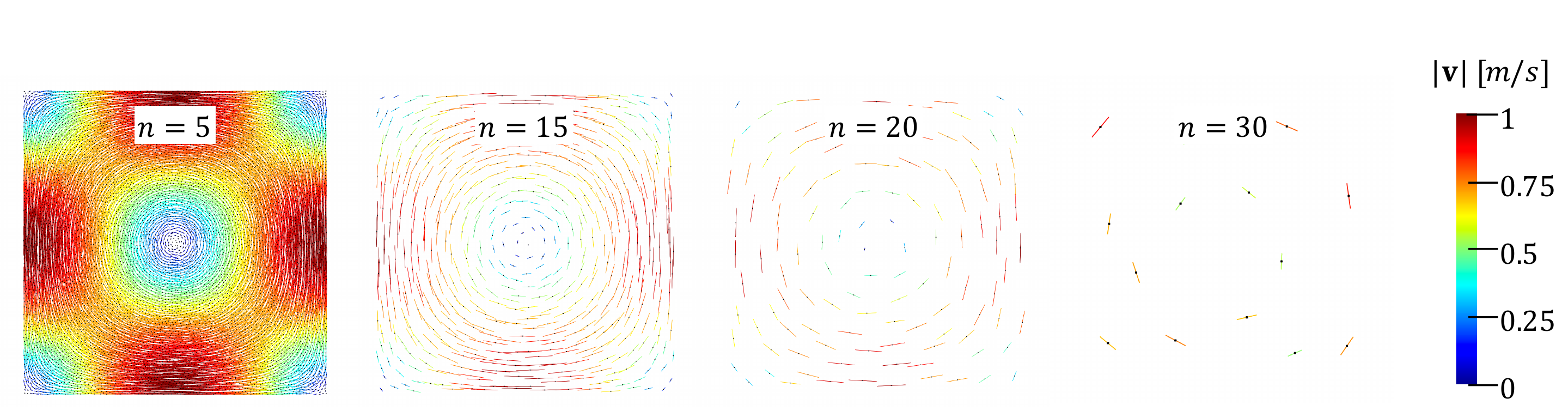}{Particle layout during the resolution coarsening with our derefinement technique. $n$ shows the number of derefinement steps performed from the initial grid.}
\customfigure{0.4}{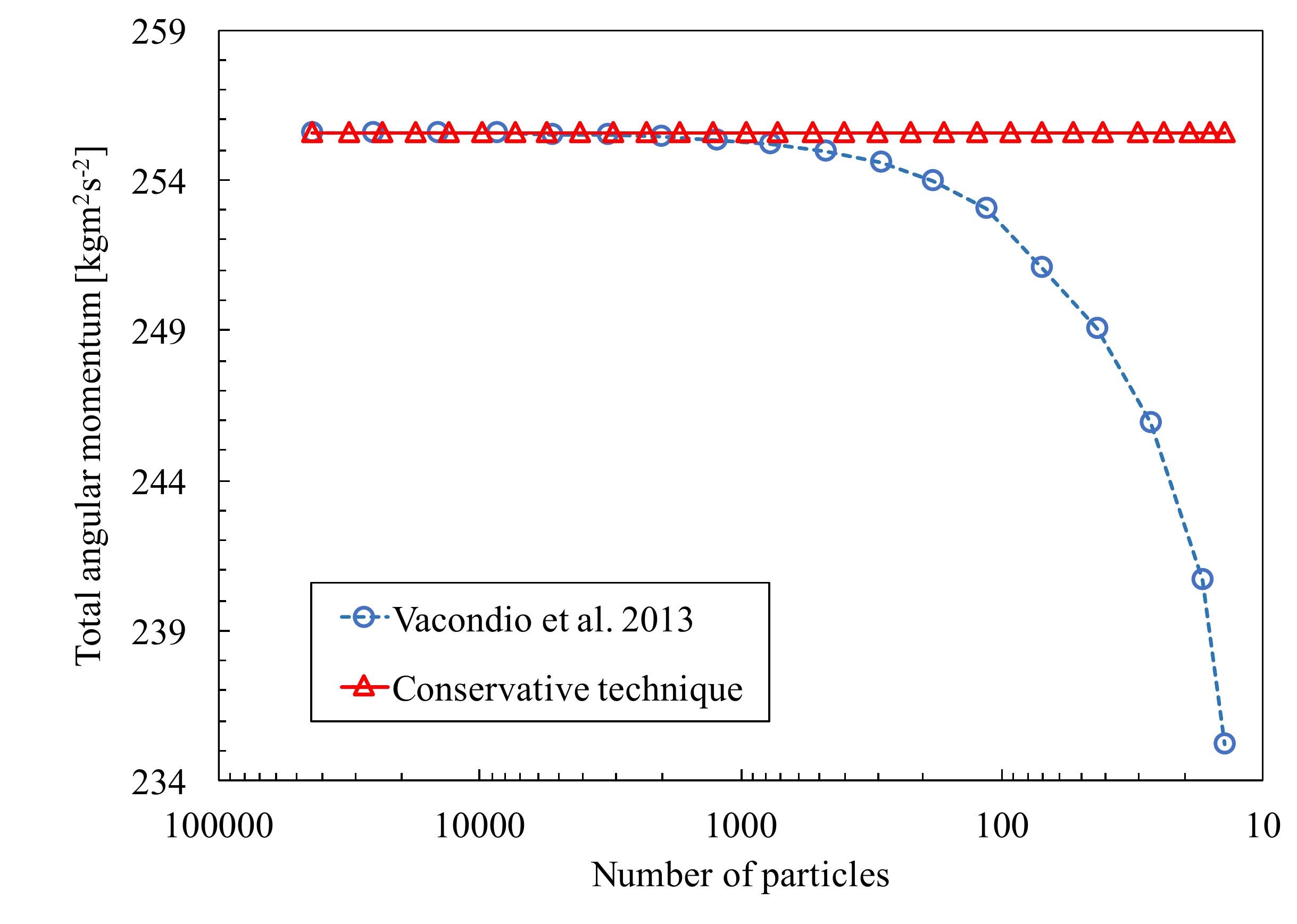}{Total angular momentum of the vortex as a function of the number of particles while sequentually merging pairs (blue line with circles) and triplets (red line with triangles).}

\revised{\subsection{Taylor-Green vortex decay}
In this section, a two-dimensional Taylor-Green vortex pattern is simulated in an origin-centered periodic box with unit edge length. Similarly to the frozen Taylor-Green example in the previous section, the initial velocity field is given as
\begin{equation} \label{eq:flowing_velocity_field}
\textbf{v}=v_0\begin{bmatrix}
           sin(2\pi x)cos(2\pi y) \\
           -cos(2\pi x)sin(2\pi y)
         \end{bmatrix},
\end{equation}
resulting in four vortices in the computational domain, where $v_0=1$. The simulations were performed using three different particle configurations yielding a uniformly coarse ($L/dx=50$) and uniformly fine resolutions ($L/dx=100$) and a multi-resolution case with dynamic particle splitting and merging in a pre-defined region shown in Figure \ref{fig:TG_init}.
\customfigure{0.7}{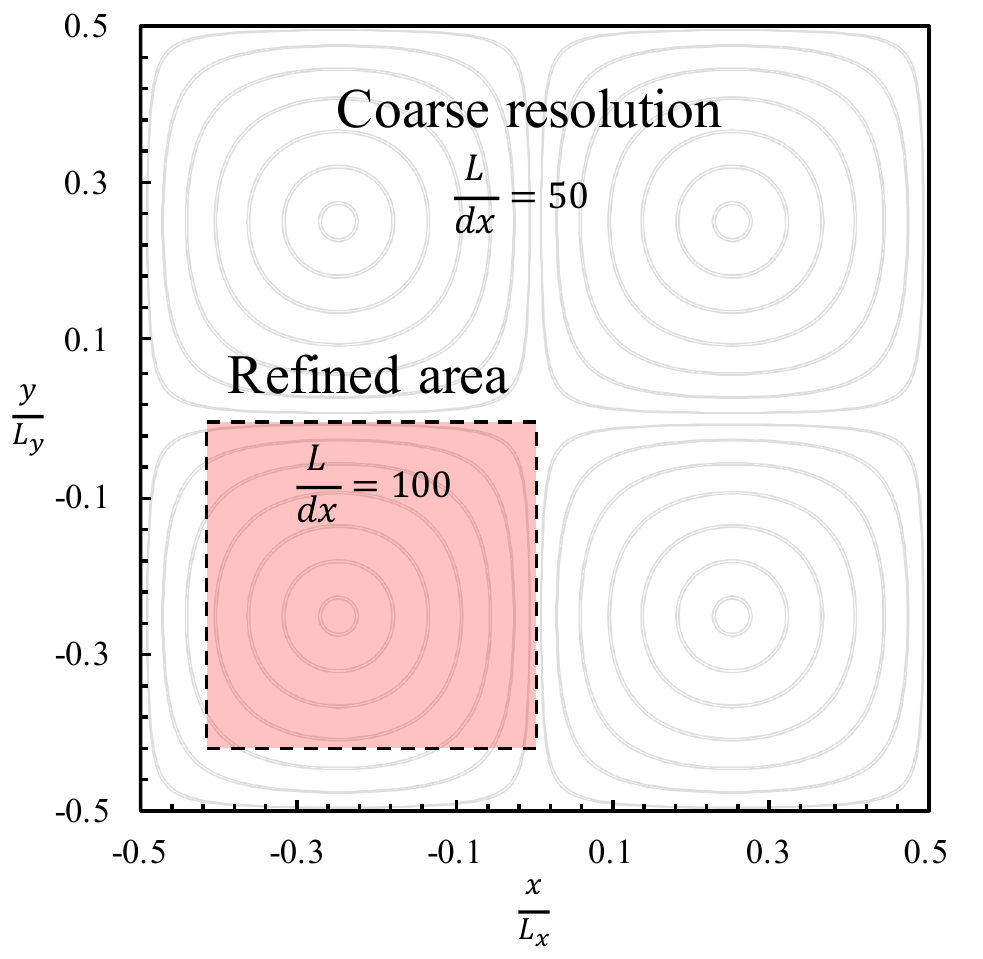}{\revised{Two-dimensional Taylor-Green vortex pattern in periodic box using dynamic particle refinement in the marked region. Gray lines show the streamlines corresponding to the initial velocity field with the four vortices.}}
The particle splitting parameters were chosen so that the refined area meet the resolution of the uniformly fine configuration. Hence, the number of daughter particles on splitting has been $n_d=4$, $\epsilon=0.3$, $\alpha=0.5$, and similarly to \cite{Barcarolo2014}, their properties concerning the mass and smoothing ratio were set to be identical. According to (\ref{eq:flowing_velocity_field}), the maximum velocity is $v_{max}=1$ m/s, which together with the density $\rho=1000$ kg/m$^3$ and kinematic viscosity $\nu=0.005$ m$^2$/s results in Re$=v_{max}L/\nu=200$. Due to the model of the current test case implies physical viscosity, the artificial diffusion term has been neglected in the momentum equation (\ref{eq:whole_system}).
\customfigure{0.45}{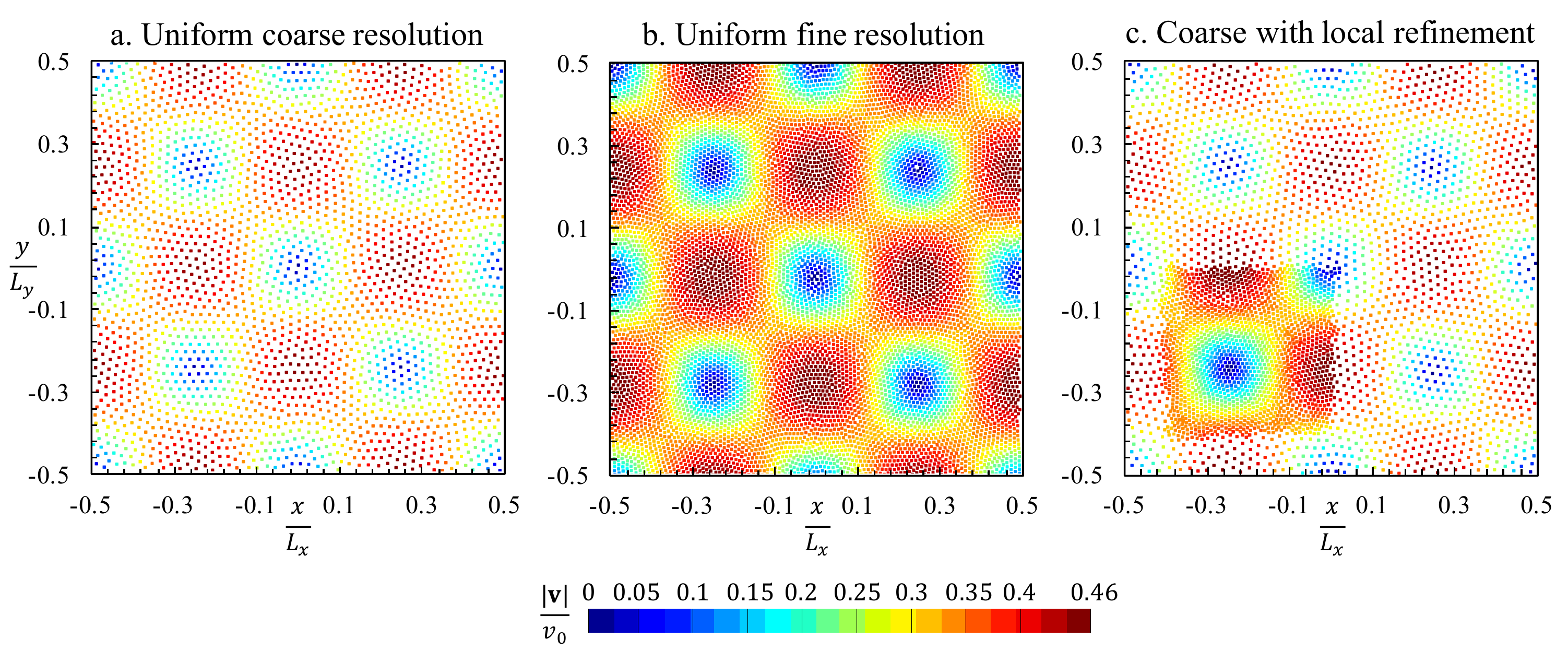}{\revised{The simulation result of the Taylor-Green vortex decay at time instant $\frac{tv_0}{L}=2$using \textit{a.} uniform coarse, \textit{b.} locally refined, and \textit{c.} uniform fine resolutions.}}
Running the simulations to the time instant $tv_0/L=2$ shows the results in Figure \ref{fig:TG_results}. Confirming that the presented coalescing technique does not affect the flow significantly, the checker-board pattern has preserved its shape in all three cases without significant distorsions. To compare the accuracy of the results, the normalized total kinetic energy $E/E_0$ was computed for each case in every instant using
\begin{equation}
E=\frac{1}{2}\sum_i^N{m_i\vert\textbf{v}_i\vert^2},
\end{equation}
where $m$ is the particle mass and $E_0$ is the kinetic energy of the initial flow. The kinetic energy decay is shown in Figure \ref{fig:TG_energy}. Although the three curves present \revisedminor{accurate evolutions compared to the analytical solution given by Kr\"uger et al. \cite{KrugerVR2010}}, the local refinement made a \revisedminor{significant} improvement over the coarse resolution case, which together with the preserved vortex pattern verifies that no significant adverse effects had been produced by our coalescing technique in the system.
\customfigure{0.45}{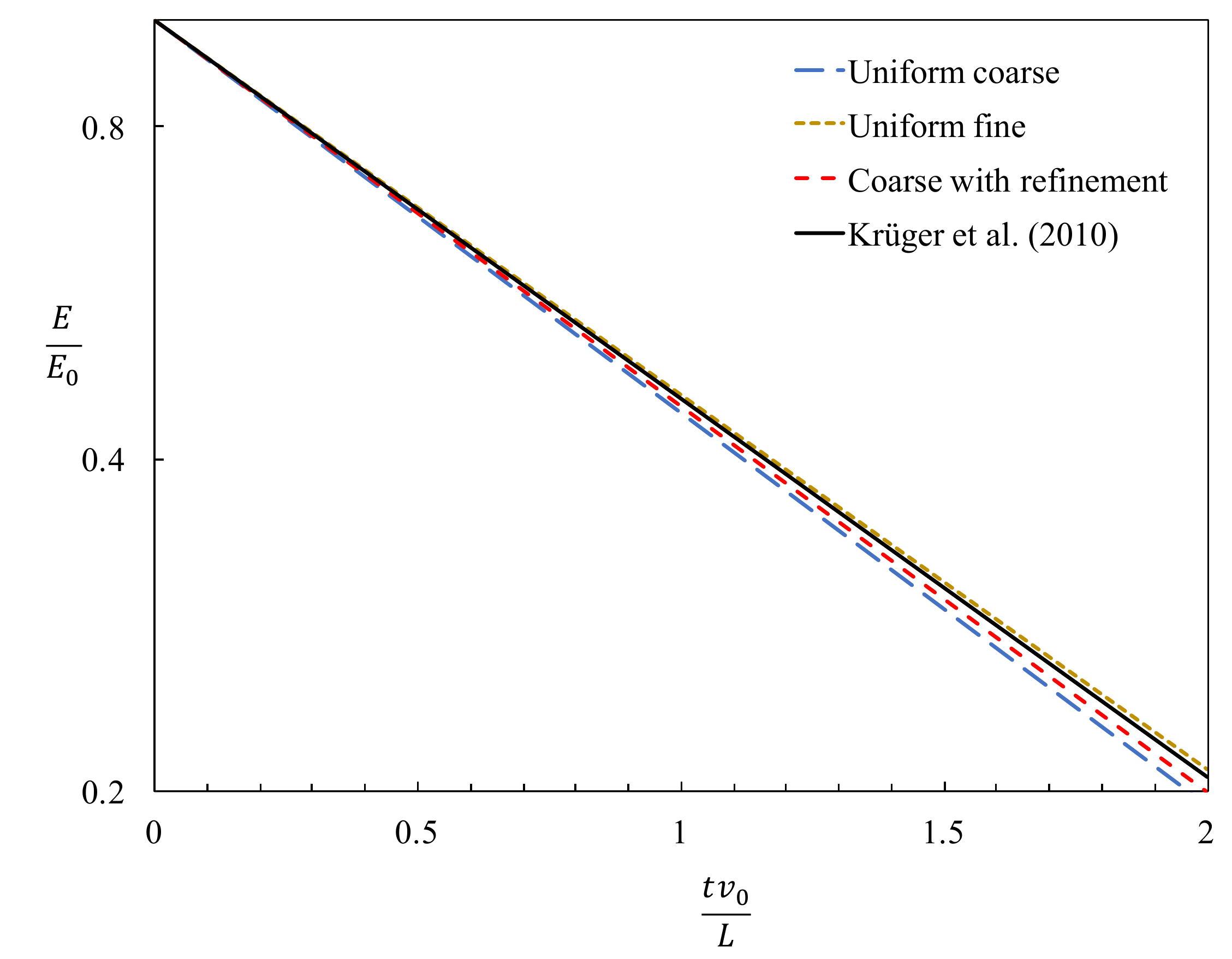}{\revised{Total kinetic energy in case of the three different configurations.}}
}
\subsection{Two-dimensional dam break}
In the third test case, the applicability of the proposed method is demonstrated by a two-dimensional dam break simulation. The applied initial geometrical layout shown in Figure \ref{fig:dambreak_draw} \revisedminor{is identical with the case experimentally investigated by Lobovsk\'y et al. \cite{Lobovsky2014}}. Using spatially fixed fluid particles, the walls of the tank were built up as a uniform grid. For the spatially varying resolution, a rectangular region with higher desired resolution was considered in the bottom right corner. \revised{Similarly to the boundary particles, fluid particles entering this region had been split up to $n_d=4$ refined or daughter particles following the pattern presented in \cite{Feldman2007}. The separation parameter and smoothing ratio were set in case of both particle types to $\epsilon=0.4$ and $\alpha=1/\sqrt{n_d}=0.5$ respectively. Also, during the replacement, the pattern of the small particles is rotated randomly in the fluid, but kept constant to form a uniform grid in the rigid wall.} According to the merging algorithm, once a refined particle leaves and moves apart from the region farther than the initial coarse interparticle distance $dx$, it is marked as a candidate for coalescing and potentially merged with two of the closest particles in the neighborhood. It is important to note that new particles do not participate in further coalescing in the same time step. During the sequence of the derefinement steps, small particles gradually restore the original coarse spatial resolution. Coloring the particles according to their pressure values, \revised{six} different time instants of the simulation results with the locally increased resolution are visible in Figure \ref{fig:dambreak_results}.

Besides the computation with the dynamic resolution, two further simulations were performed using uniformly coarse \revised{($H=81dx$)} and fine resolutions \revised{($H=161dx$)} as well. For comparison, the time series of the pressure had been evaluated at the probe $p_1$ on the right side of the tank marked with a \revised{yellow} dot in Figure \ref{fig:dambreak_draw}. To eliminate non-physical numerical oscillations, a fixed size ($\Delta=0.02$ s) renormalized Gaussian low-pass filter was applied on the pressure time series for each simulation result. The pressure probe time series of the three cases are shown in Figure \ref{fig:dambreak_pressure_series}.

\customfigure{0.4}{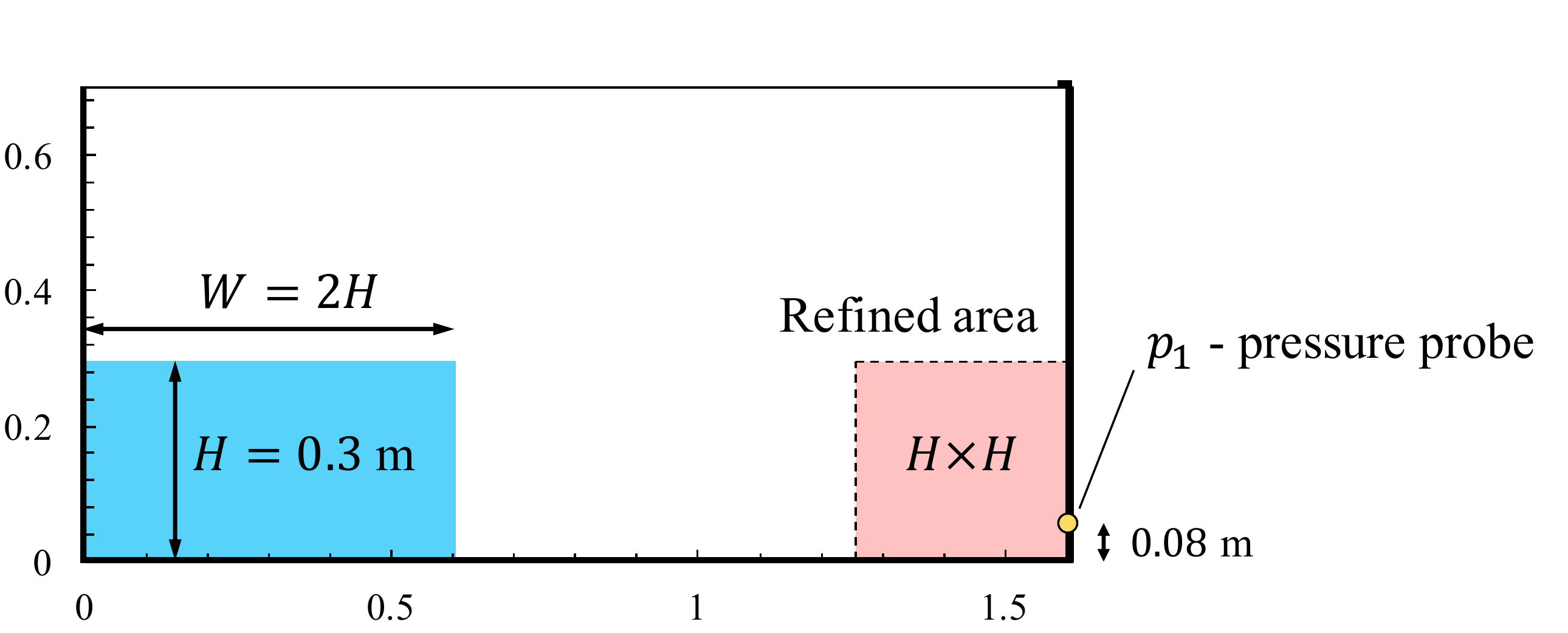}{Initial condition of the two-dimensional dam break problem. High-resolution region is visualized with the light red rectangle in the right corner, while the \revised{yellow} dot marks a pressure probe. The initial interparticle distance is denoted by $dx$.}
\customfigure{0.4}{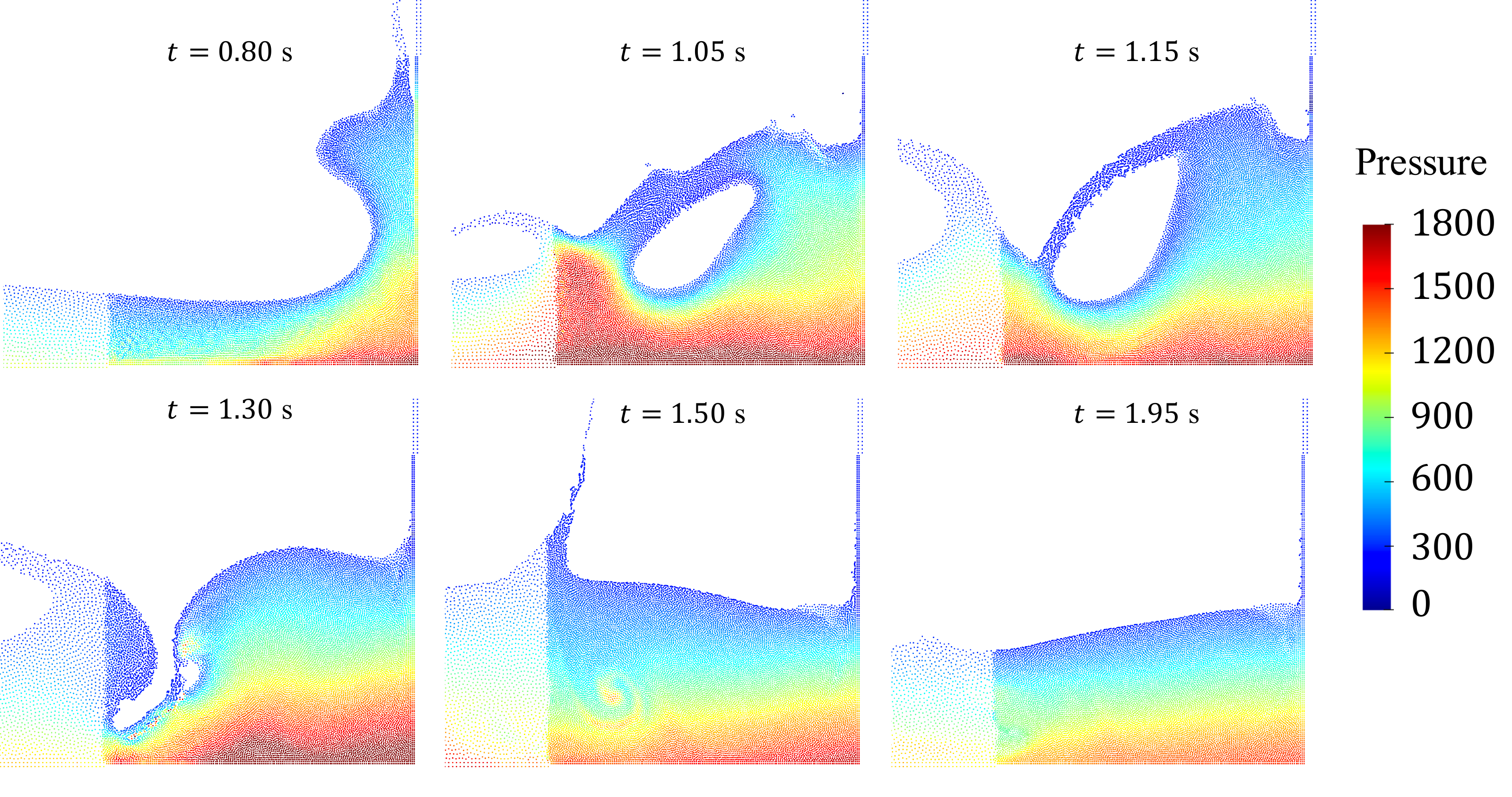}{\revised{Different time instants of the dam break simulation with dynamic splitting and conservative merging. The figure shows the bottom right corner incorporating the zone with the local refinement.}}
\customfigure{0.4}{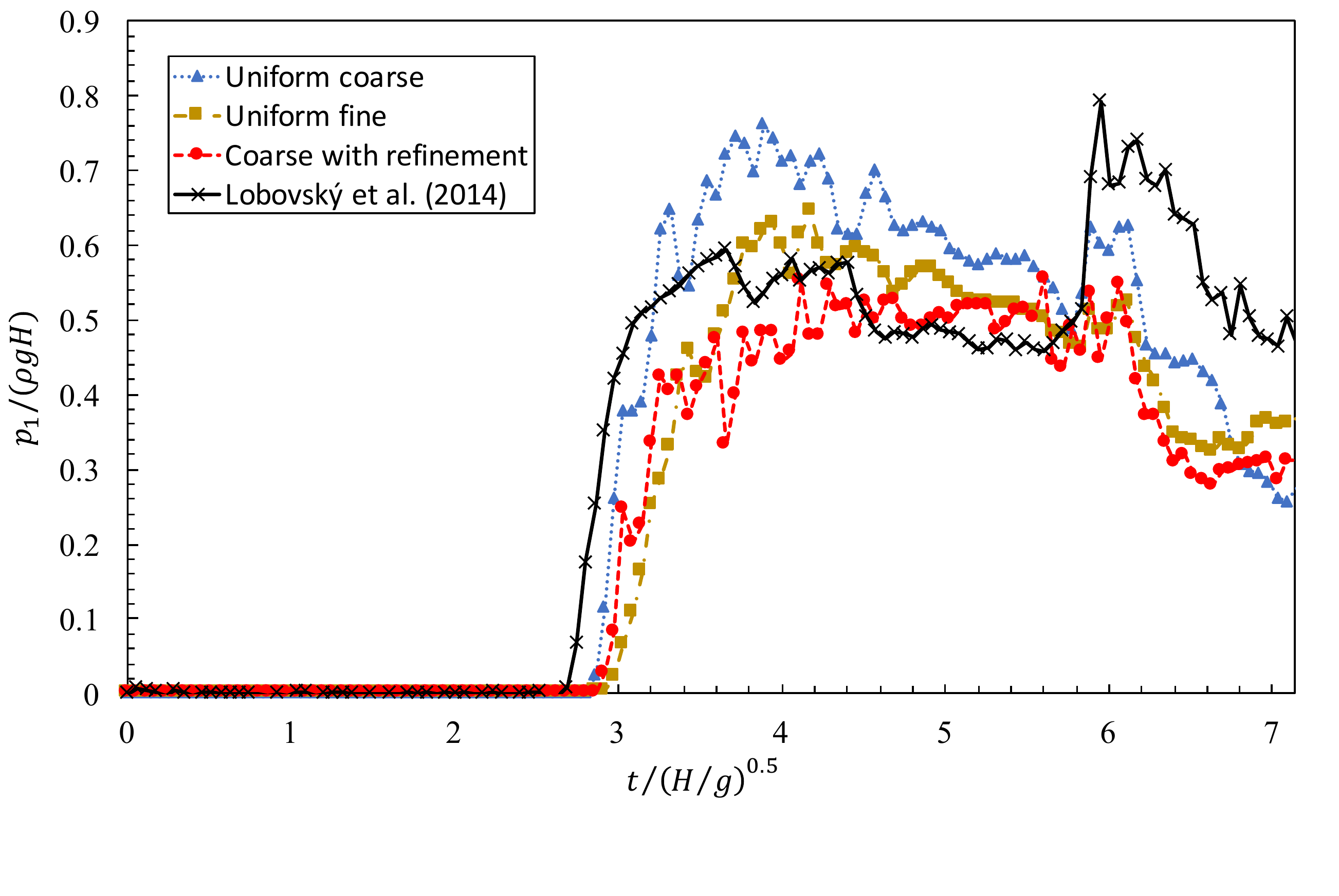}{Dam break pressure series for different resolutions (blue solid line: uniform coarse, yellow dotted line: uniform fine, red dashed line: coarse with lical refinement) evaluated at the $p_1$ probe. $a.$ pressure time series at $p_1$, $b.$ Quadratic deviation of the uniform coarse and multi-resolution cases from the uniform fine resolution.}

\section{Implementation}
The proposed conservative model and the simulation cases in the present work have been implemented and run using Nauticle \revised{introduced} in \cite{HavasiToth2019} \revised{by Tóth}, the general purpose particle-based parallel simulation tool facilitating the application and development of meshless numerical methods.
\section{Conclusion}
A novel coalescing technique with exact angular momentum conservation is introduced. The method can be considered as an extension of the former pairwise merging techniques, however, some nontrivial steps are required. During the derefinement of high resolution particles, former merging techniques replace a pair of particles with a single one inevitably canceling out local rotational motion and consequently the angular momentum. In this work, to preserve angular momentum, additional degrees of freedom for the coalesced layout are achieved by replacing three particles instead of two with a pair of identical co-rotating particles. The smoothing radius of the coalesced particles is obtained with the inverse of the Gaussian kernel so that the density error is eliminated in the local center of mass of the particles.

The proposed method has been verified through a frozen Taylor-Green vortex example with several consecutive derefinement steps from high to low resolutions. It has been shown, that pairwise techniques suffer from significant loss of angular momentum when the resolution of the vortex becomes low. Moreover, to demonstrate the efficiency of the new particle replacement scheme two-dimensional multi-resolution Taylor-Green and dam break simulations have been compared to results with uniformly coarse and fine resolutions \revisedminor{as well as with analytical and experimental data}.

Since the proposed enhanchement requires only a few additional computationally cheap steps compared to pairwise methods, its application is feasible in most circumstances \revised{and provides a more accurate particle merging technique}.

\revisedminor{The extension of the introduced coalescing technique to thee dimensions is also possible but probably not completely straighforward. In case of a three dimensional computation, the local particle layout has 9 and 6 degrees of freedom, before and after the coalescing respectively, which is still sufficient to preserve the angular momentum of the local configuration. Therefore there is no need to change the number of involved particles either in the initial or the merged layout. Moreover, since any three particles occupy a common plane in higher dimensions as well as in two dimensions, the presented replacement process is expected to be a proper choice. However, some difficulties might arise in cases when the local angular momentum vector is nearly parallel with the line between the merged particles. Although it is improbable for those to be exaclty parallel, the conservation of the angular momentum would result unphysically high local velocity magnitudes of the new particles.}

\revisedminor{
\section*{Acknowledgement}
The research reported in this paper was supported by the Higher Education Excellence Program of the Ministry of Human Capacities in the frame of Water Science \& Disaster Prevention research area of Budapest University of Technology and Economics (BME FIKP-V\'IZ)
}

\newpage
\bibliographystyle{elsarticle-num}
\bibliography{references}

\end{document}